# IBM-HBCU Quantum Center: A model for industry-academic partnerships to advance the creation of a diverse, quantum aware workforce

Kayla B. Lee
*IBM Quantum*
Yorktown Heights, NY
0000-0002-2811-7126

Thomas A. Searles
*IBM-HBCU Quantum Center*
*Dept. of Physics & Astronomy,*
*Howard University,*
Washington DC, USA
0000-0002-0532-7884

*The IBM-HBCU Quantum Center is a first-of-a-kind collaboration between IBM and a consortium of Historically Black Colleges and Universities (HBCUs) that seeks to address the lack of Black representation and build a diverse and aware workforce in quantum information science and engineering (QISE). Key pillars of the Center are focused on 1) building community and fostering a sense of belonging, 2) strengthening relationships internally and with the broader quantum community, and 3) providing funding to support undergraduate, graduate, and faculty research at HBCUs. As a part of the program, students and faculty are invited to participate in grant development workshops, a QISE invited speaker series, community hack-a-thons, and other opportunities to build competencies in the growing field of QISE. Since its launch, the IBM-HBCU Quantum Center has engaged a community of over 400 students, faculty, and researchers and will continue to establish a research presence in QISE and increase opportunities for research and workforce development.*

## I. Introduction

With the emergence of quantum computing and other quantum-based technologies in government, industry, and academia, there is a new opportunity to ensure that the future workforce reflects the diversity of the United States population more generally. In 2019, the Quantum Economic Development Consortium commissioned a survey [1] of start-ups and Fortune 500 companies in the quantum industry asking for the desired education, skill set, and experience of new hires in quantum information science and engineering (QISE). A majority of respondents named those with a PhD in Physics as the most sought-after candidates. When coupled with the fact that there are only 2-3 PhDs awarded to Black people in America each year [2], one can easily infer that within this new burgeoning industry, there is a pre-existing structure that mirrors more of the same with respect to lack of diversity and lack of opportunity for Black students. Further, the numbers become even more disparaging when one recognizes that less than 150 physics PhDs have ever been awarded to Black women in the United States. [3].

The report of the AIP National Task Force to Elevate African American Representation in Undergraduate Physics & Astronomy (TEAM-UP), titled **The Time Is Now: Systemic Changes to Increase African Americans with Bachelor's Degrees in Physics and Astronomy** [4], highlighted the well-known problem of the past 20 years: all other demographics in Physics are having record increases in the number of undergraduates with BS degrees, however "Black undergraduates" were the only group that showed a steady decline. The report attributes the success or failure of Black students in physics to factors such as a sense of belonging, physics identity, academic support, financial support, and leadership and structures. In addition, the report gave recommendations on how to address each factor. Of note, partnering with Historically Black Colleges and Universities (HBCUs) will be key in changing the future landscape.

HBCUs have a track record as a disproportionate producer of African Americans with PhDs in Physical Sciences and Engineering [Table 1, 2]. Notably, 6 out of 10 of the top baccalaureate institutions of Black science and engineering doctorate recipients are HBCUs. HBCUs have an outsized impact when it comes to graduating Black STEM majors. While HBCUs graduate 9% of the United States' Black undergraduates, 28% all Black physical science majors and 17% of all Black engineers receive their undergraduate degrees from HBCUs. This success comes despite years of underfunding from state and federal funding agencies, which when compounded with the financial vulnerability of their student bodies limits HBCU's access to extensive, cutting edge research infrastructure [5, 6]. These numbers, along with the recent TEAM-UP report highlight how HBCUs are successful at supporting students and their increasing importance in physics education.

In this article, we highlight many of the challenges in increasing diversity in the field of QISE and how the IBM-HBCU Quantum Center plans to address them by 1) building a sense of community, 2) providing academic and research support, and 3) strengthening relationships with the quantum community. In Section II, we give an overview of the IBM-HBCU Quantum Center and key goals. In Section III, we highlight identified key programmatic objectives for the first year. In Section IV, we discuss how we are measuring success and data to describe our current engagement. Finally, we discuss future plans for the program and how this applies to not only HBCUs, but other research and education programs that are interested in entering the emerging field of QISE.

TABLE I. BACCALAUREATE INSTITUTIONS AND NUMBER OF BLACK OR AFRICAN AMERICAN S&E DOCTORATE RECIPIENTS

| Baccalaureate Institution | Number of Black or African American S&E doctorate recipients, by science and engineering | Institution Size Category* (2015-19) | HBCU (Yes/No) |
|---|---|---|---|
| Spelman College | 113 | 2 | Yes |
| Howard University | 112 | 3 | Yes |
| University of Maryland-Baltimore County | 73 | 4 | No |
| University of North Carolina at Chapel Hill | 68 | 5 | No |
| Hampton University | 64 | 2 | Yes |
| North Carolina A&T State University | 62 | 4 | Yes |
| University of Maryland, College Park | 62 | 5 | No |
| Morehouse College | 61 | 2 | Yes |
| University of Florida | 61 | 5 | No |
| Florida A&M University | 58 | 3 | Yes |

## II. MODEL

The IBM-HBCU Quantum Center is a multi-year investment from IBM to support a consortium of HBCUs seeking to build a diverse and aware workforce in QISE. IBM is providing access to IBM Quantum computers via the cloud, educational support for students learning to use the Qiskit open-source software development framework [7], and funding for undergraduate, graduate and faculty research, along with opportunities for joint research collaborations.

The Center's mission is to educate, foster collaboration of joint academic-industry research, and ultimately, create a more diverse workforce for students in all STEM fields. The Center's members collaborate across their respective institutions and are building regional interactions to strengthen both faculty and student engagement. The Center is currently arranged into 5 geographical regions (Mid-Atlantic, Virginia, Carolinas, Southeastern, Southwestern) [Figure 1], each with their own focus on various quantum-based technologies, including: quantum computing, quantum simulation, quantum hardware, quantum sensing, quantum networking, and novel qubit design. In alignment with the expanding field of QISE, the Center is working with students and faculty from all STEM disciplines, with a focus on computer science, mathematics, engineering, and physics.

The design of the IBM-HBCU Quantum Center and programs were created collaboratively with HBCU faculty and

Funding source: IBM-HBCU Quantum Center

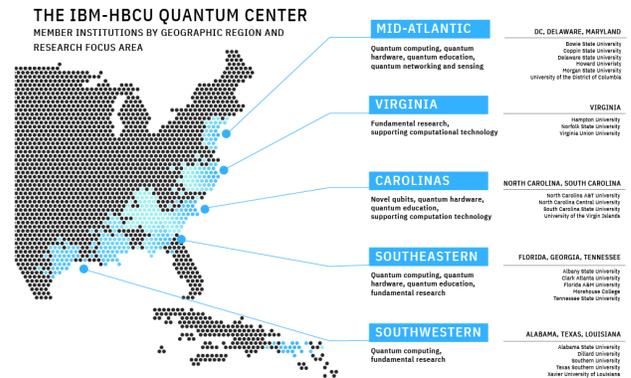

Fig. 1. IBM-HBCU Quantum Center Member Institutions by geographic region and research focus area. Member Institutions listed in alphabetical order: Alabama State University, Albany State University, Bowie State University, Clark Atlanta University, Coppin State University, Delaware State University, Dillard University, Florida A&M University, Hampton University, Howard Univeristy, Morehouse College, Morgan State University, Norfolk State University, North Carolina A&T Univeristy, North Carolina Central University, South Carolina State University, Southern University, Tennessee State University, Texas Southern University, University of District of Columbia, University of Virgin Islands, Virginia Union University, and Xavier University of Louisiana.

IBM Quantum, along with discussions from national labs and other funding agencies. The output of this work led to 3 main goals. These key pillars of the IBM-HBCU Quantum Center align to AIP TEAM-UP's recommendations and form the foundation of the program: [Figure 2]

### A. Build community and foster a sense of belonging

Fostering a sense of belonging is essential for African American student success within the context of STEM [4]. Students must perceive themselves and be perceived by others as future scientists and engineers in order to persist in QISE fields [4]. Of note, faculty, peers, and professional societies can all contribute to building students', and likely faculty participants, sense of belonging and physics identities.

As a solution to this challenge, core to the mission of the IBM-HBCU Quantum Center is building community and fostering a sense of belonging. This can be uniquely achieved by bringing together students and faculty at HBCUs to create opportunities to network, connect students with like-minded peers, and to expand the existing scope of research topics and interests. Establishing these connections will build an undergraduate community that will continue to grow throughout their QISE career. Measurable ways of meeting this goal include metrics on the number of people in the community, number of collaborative research projects, and surveying the community to better understand their experiences.

### B. Provide undergraduate and graduate research opportunities

Research experiences as an undergraduate are widely recognized as beneficial to STEM learning outcomes and are often large factors in a student's success in STEM. These experiences are associated with an increase in STEM persistence, scientific understanding, identity as a scientist, and research self-efficacy and ultimately their ability to be accepted to graduate school in STEM [8, 9]. A requirement of strong undergraduate research is access to infrastructure, even if it is not at the student's institution. However, as pointed out by Branson and Woodson [10], this access to infrastructure has been a sticking point of physics programs at HBCUs since the 1940s, where the majority of Black physics students pursue their undergraduate physics degrees [4].

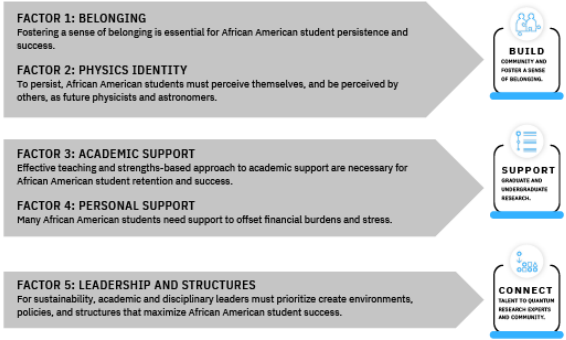

Fig. 2. Alignment to TEAM-UP and the American Institute of Physics systemic changes to increase African American with Bachelor's Degrees in Physics and Astronomy recommendations.

As a solution to this problem, we aim to support year-round research fellowships for 30 undergraduate trainees. As previously mentioned, research experiences for undergraduates (REUs) are often large factors in a student's success in QISE-related disciplines [11]. Currently, REUs, especially programs with an emphasis on recruitment for Black students, are limited to the summer months, which do not provide an opportunity to establish research continuity. By creating a year-round training program for students, which will increase their skills in the field and position them for future opportunities. Furthermore, in the context of the *AIP Time is Now* report, we are directly addressing one of the factors assigned to the decline in Black students retained in Physics [4]. Ways of measuring this goal, will be number of students trained, number of fellowships awarded, retention and graduate rates among trainees, and number of graduate school application acceptances.

*C. Establish connections to the quantum research community.*

Until the formation of the IBM-HBCU Quantum Center, the participation of HBCUs in the national quantum research community was extremely limited. In fact, from the initial total investment of 625M USD dollars into National Quantum Information Science Centers (both NSF and Department of Energy), Howard University received 1M USD, while no other funds were awarded to other HBCUs.

However, through leveraging both new and existing collaborations in QISE, the IBM-HBCU Quantum Center hopes to increase the participation of HBCU faculty and students in the larger QISE research community. As reflected in Fig. 1, the IBM-HBCU Quantum Center reaches many places that the current QIS National Center network does not reach. Therefore, it is mutually beneficial and natural to forge collaborations amongst the IBM-HBCU Quantum Center and other national research efforts. Measurable ways of gauging the goal is through participation of Center membership in quantum specific workshops/ conferences/ schools, publications in QISE-related journals, participation in visiting faculty programs, and proposals written collaboratively with National QISE Centers.

### III. Key Objectives

In order to accomplish the goals of the IBM-HBCU Quantum Center, there have been several key objectives identified to build the infrastructure for a sustainable program in QISE. These objectives include:

*A. Creating pathways to engage faculty and students in QISE*

As the field of QISE continues to be interdisciplinary, there are often questions about the field, understanding potential career pathways, applications, and skills from both students and faculty. A key feature of the IBM-HBCU Quantum Center is creating pathways to engage students and faculty from a wide range of backgrounds. Examples of how this has been accomplished include:

- IBM Quantum hosted an event for HBCU students to learn more about the budding field of Quantum computing and introduce them to internship opportunities at IBM. Using a combination of student and professor outreach, over 200 students representing 30 HBCUs attended the event. Of note, this exceeded the IBM average for participation in HBCU targeted events and includes HBCU participation outside of the current Center membership.

- Faculty of the IBM-HBCU Quantum Center are invited to a monthly invited speaker series to network and learn about ongoing research projects in the field. Drawing connections to their existing research interests and the active projects in the field can help identify future projects and potential collaborators. To date, speakers have been invited from each of the NSF Supported Quantum Leap Challenge Institutes, the Department of Energy QIS Research Centers and QIS specific national efforts in government such as NASA Goddard's quantum communications team, the Naval Research Lab and the Army Research Lab.

*B. Building research portfolio in QISE with seed grants and faculty awards*

As the government and other funding agencies continue to increase their investment in QISE, it is critical to start establishing a portfolio of research projects that map to key challenges in the field.

- As a part of this collaboration, IBM Quantum is supporting research projects in key QISE areas like quantum education [12], quantum machine learning [13, 14], quantum materials and hardware, and supporting technologies. These projects will have a faculty focal at a Member Institution in the Center with a goal of engaging undergraduate and/or graduate students in the research. The impact of these projects is two-fold: 1) it will increase opportunities for faculty research and 2) also provide students with necessary skills to establish themselves in the field.

- Notably, IBM Quantum has also partnered with other organizations to expand its reach. A collaboration between IBM and The International Society for Optics and Photonics (SPIE) led to The IBM-SPIE HBCU Faculty Accelerator Award in Quantum Optics and Photonics, an award to promote research and education in quantum optics and photonics within the Center's Member Institutions. The award may support students, postdoctoral researcher stipends, travel, conference registration, equipment, materials and supplies, and faculty summer salary.

*C. Establishing networking, recruitment, and advocacy opportunities with the quantum community.*

A key goal of the Center is the idea of 'inclusion from the start,' which will help address some of the existing disparities in the field. It is essential that both students and faculty are included in opportunities as the field matures.

- IBM Quantum is hosting a series of networking events to establish relationships between students and scientists/engineers. The purpose of the sessions are to: 1) inform students about the skills needed to be successful in internships in quantum computing, 2) encourage students to apply to internship programs, and 3) create opportunities to network with the teams that are looking for talented students.

- Since its launch, faculty and students within the Center have been invited to participate in various events, workshops, and collaborative proposals. By building a bridge between industry and other academic institutions, the Center will continue to expand its reach and ensure that both students and faculty can be active participants in the field as it matures.

## IV. Measuring Success

The IBM-HBCU Quantum Center was launched in September 2020, with the goal of increasing Black participation in quantum computing. Broadly, measuring impact includes answering these key questions:

- **Student engagement**: How can we make more HBCU students aware of opportunities in QISE and excite and engage students who express interest in QISE?

- **Talent development**: What steps do we need to take to train HBCU students to be most impactful within QISE fields and careers?

- **Workforce development**: How were HBCU students' career trajectories impacted by these experiences?

- **Research capacity**: How do we strengthen research efforts and build sustainable programs?

Within the first year, there is a strong focus on increasing participation and engagement within the Center with faculty, students, and external collaborators. Industry-academic partnerships can accelerate research and development, while also creating robust talent pipelines. This impactful relationship between academic institutions and industry also allows us to use a combination of metrics to measure success. Borrowing from traditional funnel analysis, common in marketing and growth analytics, we have mapped key metrics to a growth funnel. These metrics are 1) reach, which includes social media engagement and interactions, 2) engagement or the number of individuals attending events, engaged in online communities, and that have opted into the community, and 3) dedicated or number of users on IBM Quantum systems that have listed an HBCU as their institution affiliation.

As of the end of March 2021, the Center has been featured in over 50 news and media articles, garnering over 4000 engagements across social media platforms. In addition, when accounting for event attendance and individuals that have opted into the community, there are a total of 441 unique faculty, students, and researchers. Finally, since targeted engagement with HBCUs, registrations on IBM Quantum systems has almost doubled over the past year. These numbers highlight how initial outreach and engagement is impacting key metrics.

Moving forward, we will continue to evaluate the success of the program across multiple dimensions including: paper publications, joint research collaborations, joint research proposals applied for and awarded, students placed in internships and jobs, along with more experimental methods to increase the number of people engaged in the community.

## V. Outlook and Conclusion

The ultimate impact of the IBM-HBCU Quantum Center will be creating a foundation so that HBCUs and Black scientists can be active participants with ownership in the emerging QISE-field. This must be done in a way that ensures quantum, a relatively new field, gets built with a fundamental understanding that Black scientists have a valuable role to play within it.

As the Center continues to develop, we are measuring success on a number of metrics, including student engagement, talent and workforce development, and research capacity. Further, as the Center is funded for 5-years, we will be able to perform longer term studies on the recruitment and retention of minoritized students in interdisciplinary and emerging fields. Although the Center is primarily focused on research, there are also opportunities in the future to explore engineering and science education studies in particular with respect to infusion of quantum information into traditional STEM undergraduate curricula and the use of cloud resources in undergraduate STEM education. We hope to apply these best practices as we build the quantum workforce, especially at community colleges and undergraduate and minority-serving institutions, which all serve traditionally underrepresented communities in STEM.

## VI. Acknowledgment

T. A. S. would like to thank the IBM-HBCU Quantum Center for financial support. The views expressed are those of the authors and do not reflect official policy or position of IBM Quantum.